\documentclass[useAMS,usenatbib]{mn2e}

\usepackage[T1]{fontenc}
 \usepackage{pslatex}
\usepackage{amsmath}
\usepackage{amsfonts}
\usepackage{color}
\usepackage{url}
\usepackage{graphicx}
\usepackage{subfigure}
\usepackage{amssymb}
\usepackage{soul}
\usepackage{booktabs}
\usepackage{array}
\usepackage[utf8]{inputenc}
\inputencoding{latin1}
\inputencoding{utf8}

{\newif\ifnotend
\notendtrue
\def\veclist{ABCDEFGHIJKLMNOPQRSTUVWXYZabcdefghijklmnopqrstuvwxyz.}
\def\top#1#2.{#1}
\def\tail#1#2.{#2.}
\loop\expandafter\xdef\csname v\expandafter\top\veclist\endcsname%
{{\noexpand\bf\expandafter\top\veclist}}
\edef\veclist{\expandafter\tail\veclist}
\if\veclist.\notendfalse\fi\ifnotend\repeat}
\def\e{{\rm e}}
\def\pa{\partial}
\def\E{{\cal E}}

\mathchardef\mhyphen="2D

\title[Filamentation of FRBs in magnetar winds]{Filamentation of Fast Radio Bursts in magnetar winds}

\author[Sobacchi et al.]{Emanuele Sobacchi$^{1}$\thanks{E-mail: es3808@columbia.edu}, Yuri Lyubarsky$^2$, Andrei M. Beloborodov$^{3,4}$, Lorenzo Sironi$^1$\\
$^1$ Department of Astronomy and Columbia Astrophysics Laboratory, Columbia University, 550 West 120th Street New York, NY 10027, USA\\
$^2$ Physics Department, Ben-Gurion University, P.O.B. 653, Beer-Sheva 84105, Israel \\
$^3$ Physics Department and Columbia Astrophysics Laboratory, Columbia University, 538 West 120th Street New York, NY 10027, USA \\
$^4$ Max Planck Institute for Astrophysics, Karl-Schwarzschild-Str. 1, D-85741, Garching, Germany
}

\begin{document}

\date{}

\def\p{\partial}
\def\E{\textbf{E}}
\def\B{\textbf{B}}
\def\v{\textbf{v}}
\def\j{\textbf{j}}
\def\s{\textbf{s}}
\def\e{\textbf{e}}

\newcommand{\di}{\mathrm{d}}
\newcommand{\bfx}{\mathbf{x}}
\newcommand{\bfe}{\mathbf{e}}
\newcommand{\vlos}{\mathrm{v}_{\rm los}}
\newcommand{\Tspin}{T_{\rm s}}
\newcommand{\Tb}{T_{\rm b}}
\newcommand{\degree}{\ensuremath{^\circ}}
\newcommand{\Th}{T_{\rm h}}
\newcommand{\Tc}{T_{\rm c}}
\newcommand{\bfr}{\mathbf{r}}
\newcommand{\bfv}{\mathbf{v}}
\newcommand{\bfu}{\mathbf{u}}
\newcommand{\pc}{\,{\rm pc}}
\newcommand{\kpc}{\,{\rm kpc}}
\newcommand{\Myr}{\,{\rm Myr}}
\newcommand{\Gyr}{\,{\rm Gyr}}
\newcommand{\kms}{\,{\rm km\, s^{-1}}}
\newcommand{\de}[2]{\frac{\partial #1}{\partial {#2}}}
\newcommand{\cs}{c_{\rm s}}
\newcommand{\rb}{r_{\rm b}}
\newcommand{\rqu}{r_{\rm q}}
\newcommand{\bfOmega}{\pmb{\Omega}}
\newcommand{\bfOmegap}{\pmb{\Omega}_{\rm p}}
\newcommand{\bfXi}{\boldsymbol{\Xi}}

\maketitle

\begin{abstract}
Magnetars are the most promising progenitors of Fast Radio Bursts (FRBs). Strong radio waves propagating through the magnetar wind are subject to non-linear effects, including modulation/filamentation instabilities. We derive the dispersion relation for modulations of strong waves propagating in magnetically-dominated pair plasmas focusing on dimensionless strength parameters $a_0\lesssim 1$, and discuss implications for FRBs. As an effect of the instability, the FRB radiation intensity develops sheets perpendicular to the direction of the wind magnetic field. When the FRB front expands outside the radius where the instability ends, the radiation sheets are scattered due to diffraction. The FRB scattering timescale depends on the properties of the magnetar wind. In a cold wind, the typical scattering timescale is $\tau_{\rm sc}\sim{\rm\; \mu s-ms}$ at the frequency $\nu\sim 1{\rm\; GHz}$. The scattering timescale increases at low frequencies, with the scaling $\tau_{\rm sc}\propto\nu^{-2}$. The frequency-dependent broadening of the brightest pulse of FRB 181112 is consistent with this scaling. From the scattering timescale of the pulse, one can estimate that the wind Lorentz factor is larger than a few tens. In a warm wind, the scattering timescale can approach $\tau_{\rm sc}\sim{\rm\; ns}$. Then scattering produces a frequency modulation of the observed intensity with a large bandwidth, $\Delta\nu\sim 1/\tau_{\rm sc}\gtrsim 100{\rm\; MHz}$. Broadband frequency modulations observed in FRBs could be due to scattering in a warm magnetar wind.
\end{abstract}



\section{Introduction}

Fast Radio Bursts (FRBs) are bright flashes of millisecond duration \citep[e.g.][]{Lorimer2007, Thornton2013, Spitler+2014, Spitler2016, Petroff2016, Shannon+2018, Chime2019, Chime2019b, Chime2019a}. Magnetars are promising progenitors of FRBs. The FRB-magnetar connection, which was initially proposed on statistical grounds \citep[e.g.][]{PopovPostnov2010, PopovPostnov2013}, is strongly supported by the discovery of weak FRBs from the Galactic magnetar SGR 1935+2154 \citep[e.g.][]{Chime2020,Bochenek+2020}.

Due to the huge luminosity of FRBs, the electromagnetic field of the radio wave accelerates the electrons in the magnetar wind up to a significant fraction of the speed of light \citep[e.g.][]{LuanGoldreich2014}. Strong FRB waves can experience non-linear propagation effects, including modulation/filamentation instabilities \citep[][]{Sobacchi+2021}.\footnote{Modulation/filamentation instabilities of strong electromagnetic waves propagating in unmagnetised electron-ion plasmas are extensively studied in the field of laser-plasma interaction \citep[e.g.][]{Kruer2019}. Filamentation has been observed in numerical simulations of relativistic magnetised shocks, in which case a strong electromagnetic precursor is emitted upstream \citep[e.g.][]{Iwamoto+2017, Iwamoto+2021, BabulSironi2020, Sironi+2021}.} The instability produces a spatial modulation of the intensity of the FRB wave. As the FRB front expands, the structures generated by the instability are scattered due to diffraction, and may interfere with each other. This process leaves an imprint on the time-frequency structure of FRBs.

In our previous work \citep[][]{Sobacchi+2021}, we studied the modulation/filamentation instabilities of FRBs propagating in a weakly magnetised electron-ion plasma. Such environment may be found at large distances from the FRB progenitor. The instability develops at nearly constant electron density since the effect of the ponderomotive force is suppressed due to the large inertia of the ions. Then the dominant non-linear effect is the modification of the plasma frequency due to relativistic corrections to the effective electron mass. In regions of enhanced radiation intensity, the plasma frequency decreases because the electrons oscillate with a larger velocity in the field of the wave, and therefore have a larger effective mass. Modulations of the radiation intensity perpendicular to the direction of propagation of the wave grow because the refractive index of the plasma increases, which creates a converging lens that further enhances the radiation intensity. Modulations along the direction of propagation grow because the group velocity of the wave depends on the local radiation intensity. It turns out that the spatial scale of the modulations is shorter along the direction of propagation than in the perpendicular direction.

In this paper, we study the modulation/filamentation of FRBs propagating in a magnetar wind, which is modelled as a magnetically-dominated pair plasma. We find that the instability develops because the ponderomotive force pushes particles out of regions of enhanced radiation intensity. The refractive index of the plasma increases in the regions where the particle number density is smaller, thus creating a converging lens that further enhances the radiation intensity. Since the ponderomotive force preferentially pushes particles along the magnetic field lines, the instability produces sheets of radiation intensity perpendicular to the direction of the wind magnetic field. Consistent with previous studies focusing on unmagnetised plasmas \citep[e.g.][]{KatesKaup1989}, we do not find significant modulations of the radiation intensity along the direction of the wave propagation.\footnote{\citet[][]{ChianKennel1983} argued that strong electromagnetic waves in pair plasmas are modulated along the direction of propagation. However, these authors neglected the effect of the ponderomotive force, which is not justified in pair plasmas.}

As the FRB front expands outside the radius where the instability ends, the radiation sheets are diffracted, effectively scattering the arrival time of the FRB wave. In a cold magnetar wind, we find that the scattering timescale is $\tau_{\rm sc}\sim{\rm \mu s-ms}$. The scattering time is larger at low frequencies,\footnote{FRBs may also be scattered by some turbulent plasma screen along the line of sight. In this case, one finds $\tau_{\rm sc}\propto\nu^{-\alpha}$ with $\alpha\sim 4-4.4$ \citep[e.g.][]{LuanGoldreich2014}.} with the scaling $\tau_{\rm sc}\propto\nu^{-2}$. This scaling is consistent with the frequency-dependent broadening of the brightest pulse from FRB 181112 \citep[][]{Cho+2020}.

In a warm magnetar wind, the scattering timescale is much shorter, $\tau_{\rm sc}\sim{\rm ns}$. Then scattering produces frequency modulations with a large bandwidth, $\Delta\nu\sim 1/\tau_{\rm sc}\gtrsim 100{\rm\; MHz}$. Such broadband frequency modulations are often observed in FRBs \citep[e.g.][]{Shannon+2018, Hessels+2019, Nimmo+2021}.

The paper is organised as follows. In Section \ref{sec:basics} we briefly review some relevant properties of magnetar winds and FRBs. In Section \ref{sec:instabilities} we study the filamentation instability of FRBs. We refer the reader not interested in the technical details of the calculation to Tables \ref{table1} and \ref{table2}, where we summarise our results. In Section \ref{sec:scattering} we discuss the scattering of FRBs. In Section \ref{sec:conclusions} we conclude.

\section{Fast radio bursts in magnetar winds}
\label{sec:basics}

The magnetar wind forms outside the light cylinder, at radii $R\gtrsim R_{\rm LC}=cP/2\pi$, where $P$ is the magnetar rotational period and $c$ is the speed of light. The magnetic field strength in the wind proper frame is $B_{\rm bg}=\mu/\gamma_{\rm w}R_{\rm LC}^2R$, where $\mu$ is the magnetar magnetic dipole moment, and $\gamma_{\rm w}$ is the wind Lorentz factor. The wind magnetic field is nearly azimuthal.

The ratio of the Larmor frequency, $\omega_{\rm L}=eB_{\rm bg}/mc$, where $e$ is the electron charge $m$ is the electron mass, and the angular frequency of the FRB wave in the wind frame, $\omega_0=\pi\nu/\gamma_{\rm w}$, where $\nu$ is the observed frequency, is
\begin{equation}
\label{eq:omegaratio}
\frac{\omega_{\rm L}}{\omega_0}=2.5\times 10^{-4}\mu_{33}P_0^{-2}\nu_9^{-1}R_{15}^{-1}\;,
\end{equation}
where we have defined $\mu_{33}\equiv\mu/10^{33}{\rm\; G\; cm}^3$, $P_0\equiv P/1{\rm\; s}$, $\nu_9\equiv \nu/1{\rm\; GHz}$, and $R_{15}\equiv R/10^{15}{\rm\; cm}$. At the radii where $\omega_{\rm L}\ll\omega_0$, the electron motion is weakly affected by $B_{\rm bg}$. The electrons reach a maximum velocity of $a_0c$, where $a_0=eE/\omega_0mc$. We consider radii where $a_0\lesssim 1$, so that the electrons are sub-relativistic. The peak electric field of the wave in the wind frame, $E$, can be calculated from the isotropic equivalent of the observed FRB luminosity, $L=2c\gamma_{\rm w}^2E^2R^2$. One finds
\begin{equation}
\label{eq:a0}
a_0=2.3\times 10^{-2}\;L_{42}^{1/2}\nu_9^{-1}R_{15}^{-1}\;,
\end{equation}
where $L_{42}\equiv L/10^{42}{\rm\; erg\; s}^{-1}$. The condition $a_0\lesssim 1$ is satisfied at radii $R\gtrsim 2.3\times 10^{13}L_{42}^{1/2}\nu_9^{-1}{\rm\; cm}$. Since $\omega_{\rm L}/\omega_0\ll a_0$, the electric field of the FRB wave is larger than the wind magnetic field.

It is useful to define the wind magnetisation, $\sigma_{\rm w}$, as twice the ratio of the magnetic and rest mass energy densities of the plasma. One finds
\begin{equation}
\sigma_{\rm w}= \frac{\omega_{\rm L}^2}{\omega_{\rm P}^2} \;,
\end{equation}
where $\omega_{\rm P}=\sqrt{8\pi N_0e^2/m}$ is the plasma frequency (the particle number density is $2N_0$). We consider a magnetically-dominated wind with $\sigma_{\rm w}\gg 1$.

\section{Filamentation instability}
\label{sec:instabilities}

\subsection{Fundamental equations}

We consider an electromagnetic wave propagating through a magnetised pair plasma, with mean particle number density $2N_0$, and background magnetic field ${\bf B}_{\rm bg}$. We study the stability of slow, long-wavelength modulation of the initial wave.

The electromagnetic field of the wave can be expressed using the vector potential ${\bf A}$. We are interested in the regime where the angular frequency of the wave, $\omega_0$, is much larger than both the Larmor frequency, $\omega_{\rm L}=eB_{\rm bg}/mc$, and the plasma frequency, $\omega_{\rm P}=\sqrt{8\pi N_0e^2/m}$. When $\omega_0\gg\omega_{\rm L},\omega_{\rm P}$, the non-linear wave equation is \citep[e.g.][]{MontgomeryTidman1964, SluijterMontgomery1965, Ghosh+2021}
\begin{equation}
\label{eq:wave}
\frac{\pa^2{\bf A}}{\pa t^2}-c^2\nabla^2{\bf A}+\omega_{\rm P}^2\frac{\langle N\rangle}{N_0}\left(1-\frac{1}{2}\frac{e^2\langle A^2\rangle}{m^2c^2}\right){\bf A}=0\;,
\end{equation}
where $\langle\ldots\rangle$ denotes the average on the spatial scale of many wavelengths of the initial wave. We have defined $\langle N\rangle$ as half the total particle number density, namely $2\langle N\rangle=\langle N_{e^+}\rangle + \langle N_{e^-}\rangle$ where $\langle N_{e^+}\rangle$ and $\langle N_{e^-}\rangle$ are the positron and electron densities. To avoid a lengthy notation, below we write $N_{e^\pm}$ instead of $\langle N_{e^\pm}\rangle$.

Eq. \eqref{eq:wave} contains two non-linear terms. The term proportional to $\langle A^2\rangle$ originates from the relativistic corrections to the effective electron mass, and from the beating between the density oscillations at the frequency $2\omega_0$ and the velocity oscillations at the frequency $\omega_0$ \citep[for a detailed discussion, see Appendix A of][]{Ghosh+2021}. The term $\langle N\rangle$ describes plasma density modulations produced by the ponderomotive force.

We use the same approach that is customarily adopted to study non-linear propagation effects in unmagnetised electron-ion plasmas \citep[e.g.][]{Kruer2019}. The plasma is described using a two-fluid model. The evolution of the positron and electron number densities is described by the continuity equation
\begin{equation}
\label{eq:cont+-}
\frac{\pa N_{e^\pm}}{\pa t} + \nabla\cdot\left(N_{e^\pm}{\bf V}_{e^\pm}\right)=0\;,
\end{equation}
where ${\bf V}_{e^+}$ and ${\bf V}_{e^-}$ are the positron and electron coordinate velocities. The evolution of the velocities is described by the Euler's equation
\begin{align}
\nonumber
\frac{\pa{\bf V}_{e^\pm}}{\pa t} + & \left({\bf V}_{e^\pm}\cdot\nabla\right){\bf V}_{e^\pm} = -c_{\rm s}^2\frac{\nabla N_{e^\pm}}{N_{e^\pm}} +\\
\label{eq:euler+-}
\pm & \frac{e}{m}\left[{\bf E}+\frac{{\bf V}_{e^\pm}}{c}\times\left({\bf B}+{\bf B}_{\rm bg}\right)\right] - \frac{1}{2}\frac{e^2}{m^2}\nabla\langle A^2\rangle
\end{align}
where $c_{\rm s}$ is the thermal velocity. The last term of Eq. \eqref{eq:euler+-} is the gradient of the ponderomotive potential.

The electric field ${\bf E}$ and the magnetic field ${\bf B}$ obey Maxwell's equations
\begin{align}
\label{eq:maxwell1}
\nabla\cdot{\bf E} & = 4\pi e\left(N_{e^+}-N_{e^-}\right) \\
\label{eq:maxwell2}
\nabla\cdot{\bf B} & = 0 \\
\label{eq:maxwell3}
\nabla\times{\bf E} & = -\frac{1}{c}\frac{\pa {\bf B}}{\pa t} \\
\label{eq:maxwell4}
\nabla\times{\bf B} & = \frac{4\pi}{c}e\left(N_{e^+}{\bf V}_{e^+}-N_{e^-}{\bf V}_{e^-}\right)+\frac{1}{c}\frac{\pa{\bf E}}{\pa t}\;.
\end{align}
We remark that all the physical quantities in Eqs. \eqref{eq:cont+-}-\eqref{eq:maxwell4} describe oscillations at frequencies much smaller than $\omega_0$.

The remainder of this section is organised as follows. In Section \ref{sec:pump} we find a solution of Eqs. \eqref{eq:wave}-\eqref{eq:maxwell4} that is independent of $x$ and $y$ (such solution is called ``electromagnetic pump wave''). In Sections \ref{sec:pert} and \ref{sec:DR} we study the stability of the initial pump wave. We focus on the regime of sub-relativistic electron motion, i.e. $a_0=eA_0/mc\lesssim 1$ and $\beta_{\rm s}=c_{\rm s}/c\ll 1$.

\subsection{Electromagnetic pump wave}
\label{sec:pump}

The electromagnetic pump wave is described by the vector potential
\begin{equation}
\label{eq:apump}
{\bf A}=\frac{1}{2}{\bf A}_0\exp\left({\rm i}\omega_0 t-{\rm i}{\bf k}_0 \cdot{\bf x}\right) +{\rm c.c.} \;,
\end{equation}
where ${\bf A}_0$ is real, and ${\rm c.c.}$ indicates the complex conjugate (then Eq. \eqref{eq:apump} gives ${\bf A}={\bf A}_0\cos(\omega_0 t-{\bf k}_0 \cdot{\bf x})$). Since $\langle A^2\rangle=A_0^2/2$, the gradient of the ponderomotive potential vanishes. Then Eqs. \eqref{eq:cont+-}-\eqref{eq:maxwell4} have the straightforward solution $N_{e^\pm}=N_0$, ${\bf V}_{e^\pm}=0$, and ${\bf E}={\bf B}=0$. Substituting Eq. \eqref{eq:apump} into Eq. \eqref{eq:wave}, one finds the dispersion relation of the pump wave,
\begin{equation}
\label{eq:DRpump}
\omega_0^2 = c^2k_0^2+\omega_{\rm P}^2\left(1-\frac{1}{4}a_0^2\right) \;,
\end{equation}
where
\begin{equation}
\label{eq:a0def}
a_0=\frac{eA_0}{mc}\;.
\end{equation}
Eq. \eqref{eq:DRpump} is the classical result of \citet[][]{SluijterMontgomery1965}.

\subsection{Small perturbations}
\label{sec:pert}

Modulations with frequency $\omega$ and wave vector ${\bf k}$ of intensity of the pump wave are described by two beating wavebands with frequencies $\omega_\pm=\omega\pm\omega_0$ and wave vectors ${\bf k}_\pm={\bf k}\pm{\bf k}_0$, where $\omega^2\ll\omega_0^2$ and $k^2\ll k_0^2$. The perturbed vector potential is
\begin{align}
\nonumber
{\bf A} & = \frac{1}{2}{\bf A}_0\exp\left({\rm i}\omega_0 t-{\rm i}{\bf k}_0 \cdot{\bf x}\right) + \\
\label{eq:deltaA}
& \phantom{=}+\delta {\bf A}_+\exp\left({\rm i}\omega_+t-{\rm i}{\bf k}_+\cdot{\bf x}\right)+ \delta {\bf A}_-\exp\left({\rm i}\omega_-t-{\rm i}{\bf k}_-\cdot{\bf x}\right)+{\rm c.c.}\;,
\end{align}
where ${\bf A}_0$ and $\delta{\bf A}_\pm$ are nearly aligned. Writing ${\bf A}_0=A_0{\bf n}$ and $\delta {\bf A}_\pm=\delta A_\pm{\bf n}$, where ${\bf n}$ is a unit vector, from Eq. \eqref{eq:deltaA} one finds
\begin{equation}
\label{eq:Asq}
\langle A^2\rangle-\frac{1}{2}A_0^2=A_0\left(\delta A_+ + \delta A_-\right)\exp\left({\rm i}\omega t-{\rm i}{\bf k}\cdot{\bf x}\right)+{\rm c.c.} \;,
\end{equation}
where we have neglected quadratic terms in the perturbed quantities. The average is made on a spatial scale much longer than $k_0^{-1}$, and much shorter than $k^{-1}$.

Below we use Eqs. \eqref{eq:cont+-}-\eqref{eq:maxwell4} to calculate the density perturbation $\delta N=\delta N_{e^+}+\delta N_{e^-}$ as a function of $\delta A_+$ and $\delta A_-$. Then we substitute $\delta N$ into Eq. \eqref{eq:wave} and derive two homogeneous equations for $\delta A_+$ and $\delta A_-$. The condition that the determinant of the coefficients vanishes gives the dispersion relation.

\subsubsection{Two-fluid equations}

Substituting $N_{e^\pm}=N_0+\delta N_{e^\pm}$, ${\bf V}_{e^\pm}=\delta {\bf V}_{e^\pm}$, into Eqs. \eqref{eq:cont+-}-\eqref{eq:euler+-}, and neglecting quadratic terms in the perturbed quantities, one finds
\begin{align}
\label{eq:pertcont+-}
\frac{\pa\delta N_{e^\pm}}{\pa t} + & N_0\nabla\cdot\delta {\bf V}_{\rm e^\pm}=0\\
\label{eq:perteuler+-}
\frac{\pa\delta {\bf V}_{e^\pm}}{\pa t} = & -c_{\rm s}^2\frac{\nabla\delta N_{e^\pm}}{N_0} \pm \frac{e}{m}\left[\delta{\bf E}+\frac{\delta{\bf V}_{e^\pm}}{c}\times{\bf B}_{\rm bg}\right] - \frac{1}{2}\frac{e^2}{m^2}\nabla\langle A^2\rangle \;.
\end{align}
It is convenient to introduce new variables defined as $2\delta N=\delta N_{e^+}+\delta N_{e^-}$, $2\delta N_{\rm a}=\delta N_{e^+}-\delta N_{e^-}$, $2\delta{\bf V}=\delta{\bf V}_{e^+}+\delta{\bf V}_{e^-}$, and $2\delta{\bf V}_{\rm a}=\delta{\bf V}_{e^+}-\delta{\bf V}_{e^-}$. From Eq. \eqref{eq:pertcont+-}, one finds
\begin{align}
\label{eq:p1}
\frac{\pa\delta N}{\pa t} & +N_0\nabla\cdot\delta{\bf V}=0 \\
\label{eq:p2}
\frac{\pa\delta N_{\rm a}}{\pa t} & +N_0\nabla\cdot\delta{\bf V}_{\rm a}=0\;.
\end{align}
From Eq. \eqref{eq:perteuler+-}, one finds
\begin{align}
\label{eq:p3}
\frac{\pa\delta {\bf V}}{\pa t} & = -c_{\rm s}^2\frac{\nabla\delta N}{N_0} + \frac{e}{m}\frac{\delta{\bf V}_{\rm a}}{c}\times{\bf B}_{\rm bg} - \frac{1}{2}\frac{e^2}{m^2}\nabla\langle A^2\rangle \\
\label{eq:p4}
\frac{\pa\delta {\bf V}_{\rm a}}{\pa t} & = -c_{\rm s}^2\frac{\nabla\delta N_{\rm a}}{N_0} + \frac{e}{m}\left[\delta{\bf E}+\frac{\delta{\bf V}}{c}\times{\bf B}_{\rm bg}\right]\;.
\end{align}
Substituting ${\bf E}=\delta{\bf E}$, ${\bf B}=\delta{\bf B}$ into Eqs. \eqref{eq:maxwell1}-\eqref{eq:maxwell4}, one finds
\begin{align}
\label{eq:p5}
\nabla\cdot\delta{\bf E} & = 8\pi e\delta N_{\rm a} \\
\label{eq:p6}
\nabla\cdot\delta{\bf B} & = 0 \\
\label{eq:p7}
\nabla\times\delta{\bf E} & = -\frac{1}{c}\frac{\pa\delta{\bf B}}{\pa t} \\
\label{eq:p8}
\nabla\times\delta{\bf B} & = \frac{8\pi}{c}eN_0\delta{\bf V}_{\rm a}+\frac{1}{c}\frac{\pa\delta{\bf E}}{\pa t}\;.
\end{align}
It is convenient to introduce a system of coordinates $(x',y',z')$ so that ${\bf k}=k{\bf e}_{z'}$ and ${\bf B}_{\rm bg}=B_{\rm bg}\sin\theta{\bf e}_{y'}+B_{\rm bg}\cos\theta{\bf e}_{z'}$. Since $\nabla\langle A^2\rangle$ is directed along ${\bf k}$, the solution has the form $\delta{\bf V}=\delta V_{y'}{\bf e}_{y'}+\delta V_{z'}{\bf e}_{z'}+{\rm c.c.}$, $\delta{\bf V}_{\rm a}=\delta V_{{\rm a,}x'}{\bf e}_{x'}+{\rm c.c.}$, $\delta N_{\rm a}=0$, $\delta{\bf E}=\delta E_{x'} {\bf e}_{x'}+{\rm c.c.}$, $\delta{\bf B}=\delta B_{y'}{\bf e}_{y'}+{\rm c.c.}$ Since $\nabla\langle A^2\rangle$ is proportional to $\exp({\rm i}\omega t-{\rm i}{\bf k}\cdot{\bf x})$, all the variables depend on the coordinates as $\exp({\rm i}\omega t-{\rm i}{\bf k}\cdot{\bf x})$.

Eq. \eqref{eq:p7} gives $\delta B_{y'}=(ck/\omega)\delta E_{x'}$. Substituting $\delta B_{y'}=(ck/\omega)\delta E_{x'}$ into Eq. \eqref{eq:p8}, one finds
\begin{equation}
\label{eq:Ex}
\frac{e}{m}\delta E_{x'}={\rm i}\frac{\omega\omega_{\rm P}^2}{\omega^2-c^2k^2}\delta V_{{\rm a,}x'}\;.
\end{equation}
The $y'$ component of Eq. \eqref{eq:p3} gives
\begin{equation}
\label{eq:Vy}
\delta V_{y'}={\rm i}\frac{\omega_{\rm L}}{\omega}\cos\theta\delta V_{{\rm a},x'}\;.
\end{equation}
Using Eqs. \eqref{eq:Ex}-\eqref{eq:Vy}, from Eq. \eqref{eq:p4} one finds
\begin{equation}
\label{eq:Vz1}
\sin\theta\delta V_{z'}+{\rm i} \frac{\omega}{\omega_{\rm L}}\left(1-\frac{\omega_{\rm P}^2}{\omega^2-c^2k^2}-\frac{\omega_{\rm L}^2}{\omega^2}\cos^2\theta\right)\delta V_{{\rm a,}x'}=0\;.
\end{equation}
Since $\delta N/N_0=(k/\omega)\delta V_{z'}$, which follows from Eq. \eqref{eq:p1}, the $z'$ component of Eq. \eqref{eq:p3} gives
\begin{align}
\nonumber
{\rm i}\omega\left(1-\frac{c_{\rm s}^2k^2}{\omega^2}\right) & \delta V_{z'} -\omega_{\rm L}\sin\theta\delta V_{{\rm a,}x'}=\\
\label{eq:Vz2}
& =\frac{{\rm i}}{2}\frac{e^2}{m^2} kA_0\left(\delta A_+ +\delta A_-\right)\exp\left({\rm i}\omega t-{\rm i}{\bf k}\cdot{\bf x}\right) \;,
\end{align}
where we have used Eq. \eqref{eq:Asq} to calculate $\nabla\langle A^2\rangle$. Obtaining $\delta V_{z'}$ from Eqs. \eqref{eq:Vz1}-\eqref{eq:Vz2}, and using the fact that $\langle N\rangle/N_0-1=(k/\omega)\delta V_{z'}+{\rm c.c.}$, one eventually finds
\begin{equation}
\label{eq:N}
\frac{\langle N\rangle}{N_0}-1=\frac{Q}{2}\frac{e^2}{m^2c^2}A_0\left(\delta A_+ +\delta A_-\right)\exp\left({\rm i}\omega t-{\rm i}{\bf k}\cdot{\bf x}\right)+{\rm c.c.} \;,
\end{equation}
where
\begin{equation}
\label{eq:Q}
Q= \frac{\frac{c^2k^2}{\omega^2}\left(1-\frac{\omega_{\rm P}^2}{\omega^2-c^2k^2}-\frac{\omega_{\rm L}^2}{\omega^2}\cos^2\theta\right)}{\left(1-\frac{c_{\rm s}^2k^2}{\omega^2}\right) \left(1-\frac{\omega_{\rm P}^2}{\omega^2-c^2k^2}-\frac{\omega_{\rm L}^2}{\omega^2}\cos^2\theta\right)-\frac{\omega_{\rm L}^2}{\omega^2}\sin^2\theta} \;.
\end{equation}
Note that the density perturbation is independent of $\omega_{\rm L}$ when the ponderomotive force is aligned with the magnetic field. Indeed, for $\theta=0$ one finds $Q=c^2k^2/(\omega^2-c_{\rm s}^2k^2)$.

\subsubsection{Dispersion relation}

Substituting Eqs. \eqref{eq:deltaA}, \eqref{eq:Asq}, and \eqref{eq:N} into Eq. \eqref{eq:wave}, and neglecting quadratic terms in the perturbed quantities, one finds
\begin{equation}
\label{eq:wavepert}
S_+\exp\left({\rm i}\omega_+t-{\rm i}{\bf k}_+\cdot{\bf x}\right)+S_-\exp\left({\rm i}\omega_-t-{\rm i}{\bf k}_-\cdot{\bf x}\right)+{\rm c.c.}=0\;,
\end{equation}
where
\begin{align}
\nonumber
S_\pm= \left[\omega_\pm^2-c^2k_\pm^2\right. & -\left. \omega_{\rm P}^2\left(1-\frac{1}{4}a_0^2\right)\right]\delta A_\pm + \\
& + \frac{1}{4}\left(1-Q\right)a_0^2\omega_{\rm P}^2\left(\delta A_+ +\delta A_-\right) \;.
\end{align}
Eq. \eqref{eq:wavepert} requires $S_+=S_-=0$, which is a homogeneous system of two equations for $\delta A_+$ and $\delta A_-$. The condition that the determinant of the coefficients vanishes gives the dispersion relation. Since $\omega_\pm^2-c^2k_\pm^2-\omega_{\rm P}^2(1-a_0^2/4)=(\omega^2-c^2k^2)\pm 2(\omega_0\omega-c^2{\bf k}_0\cdot{\bf k})$, which follows from Eq. \eqref{eq:DRpump}, the dispersion relation can be presented as
\begin{align}
\nonumber
4\left(\omega_0\omega-c^2{\bf k}_0\cdot{\bf k}\right)^2 = & \left(\omega^2-c^2k^2\right)^2  +\\
\label{eq:DR}
& + \frac{1}{2}\left(1-Q\right)a_0^2\omega_{\rm P}^2\left(\omega^2-c^2 k^2\right) \;,
\end{align}
where $a_0$ and $Q$ are given by Eqs. \eqref{eq:a0def} and \eqref{eq:Q}.

\subsection{Evolution of the wavebands}
\label{sec:DR}

Below we solve the dispersion relation, Eq. \eqref{eq:DR}, and show that the wavebands grow exponentially. We are interested in a magnetically-dominated magnetar wind with $\sigma_{\rm w}=\omega_{\rm L}^2/\omega_{\rm P}^2\gg 1$. We focus on a pump wave that propagates in the direction perpendicular to the background magnetic field, as expected since the wind magnetic field is nearly azimuthal. We introduce a system of coordinates $(x,y,z)$ so that ${\bf k}_0=k_0{\bf e}_z$, ${\bf k}=k_y{\bf e}_y+k_z{\bf e}_z$, and ${\bf B}_{\rm bg}=B_{{\rm bg,}x} {\bf e}_x+B_{{\rm bg,}y} {\bf e}_y$. The cosine of the angle between the ponderomotive force (which is directed along ${\bf k}$) and the background magnetic field is $\cos^2\theta=k_y^2B_{{\rm bg,}y}^2/(k_y^2+k_z^2)(B_{{\rm bg,}x}^2+B_{{\rm bg,}y}^2)$.

\begin{table*}
\begin{center}
\begin{tabular}{cccc}
range of $a_0$ & $ck_y$ & $ck_z$ & $\Gamma$ \\ [0.5ex] 
\hline
$a_0\gg \beta_{\rm s}^2\omega_0/\omega_{\rm P}$ & $\sqrt{a_0\omega_0\omega_{\rm P}} \ll ck_y\ll a_0\beta_{\rm s}^{-1}\omega_{\rm P}$ & $\ll a_0\omega_{\rm P}$ & $a_0\omega_{\rm P}/\sqrt{2}$ \\ [0.5ex] 
$a_0\ll \beta_{\rm s}^2\omega_0/\omega_{\rm P}$ & $ck_y\simeq a_0\beta_{\rm s}^{-1}\omega_{\rm P}/2$ & $\ll a_0\omega_{\rm P}$ & $a_0^2\beta_{\rm s}^{-2} \omega_{\rm P}^2 /8\omega_0$ \\
\end{tabular}
\caption{\label{table1} Wave number in the transverse direction ($k_y$) and in the longitudinal direction ($k_z$), and growth rate ($\Gamma$) of the unstable modes (the electromagnetic pump wave is propagating along the $z$ direction). Results are the same for a weakly magnetised wind ($\sigma_{\rm w}\ll a_0^2$), and for a magnetically-dominated wind with the background magnetic field along $k_y$. The growth rate remains same order of the maximal one for $k_z\ll a_0\omega_{\rm P}/c$, while there is no instability for $k_z\gg a_0\omega_{\rm P}/c$. When $a_0\gg\beta_{\rm s}^2\omega_0/\omega_{\rm P}$, the growth rate remains same order of the maximal one for $\sqrt{a_0\omega_0\omega_{\rm P}}/c\ll k_y\ll a_0\beta_{\rm s}^{-1}\omega_{\rm P}/c$. Since $k_y\gg k_z$, the modulations are elongated in the direction of propagation of the electromagnetic pump wave.}
\end{center}
\end{table*}

\begin{table}
\begin{center}
\begin{tabular}{ccc}
$ck_y$ & $ck_z$ & $\Gamma$ \\ [0.5ex]
\hline
$a_0\omega_{\rm P}/2$ & $\ll\omega_{\rm L}$ & $a_0^2\omega_{\rm P}^2/8\omega_0$ \\
\end{tabular}
\caption{\label{table2} Same as Table \ref{table1}, but for a magnetically-dominated wind with the background magnetic field perpendicular to $k_y$ and $k_z$. The growth rate remains same order of the maximal one for $k_z\ll \omega_{\rm L}/c$, while there is no instability for $k_z\gg \omega_{\rm L}/c$.}
\end{center}
\end{table}

\subsubsection{Case $\omega^2\gg\omega_{\rm L}^2$}

It is convenient to start with the case $\omega^2\gg\omega_{\rm L}^2$, since the background magnetic field does not affect the development of the instability. Indeed, one can approximate $Q\simeq c^2k^2/(\omega^2-c_{\rm s}^2k^2)$, which is independent of $\omega_{\rm L}$.\footnote{Since $Q\gg 1$, in Eq. \eqref{eq:wave} one has $\delta\langle N\rangle/N_0 \gg (e^2/m^2c^2)\delta\langle A^2\rangle$. Then the density modulations produced by the ponderomotive force are the dominant non-linear effect leading to the exponential growth of the instability.} Since the dispersion relation depends only on $k_z$ and $k^2$, when $\omega^2\gg\omega_{\rm L}^2$ there is a rotational symmetry about the direction of propagation of the pump wave.

One can find approximate analytical solutions of Eq. \eqref{eq:DR} as follows. Far from the resonances the right hand side of Eq. \eqref{eq:DR} is small. Then the solution can be presented as $\omega= c^2k_0k_z/\omega_0+\Delta\omega$, with a small $\Delta\omega$. Substituting $\omega= c^2k_0k_z/\omega_0+\Delta\omega$, the left hand side of Eq. \eqref{eq:DR} becomes $(\omega_0\omega-c^2k_0k_z)^2 = \omega_0^2(\Delta\omega)^2$.

Now we discuss the approximation of $\omega^2-c^2k^2$ on the right hand side of Eq. \eqref{eq:DR}. Substituting $\omega= c^2k_0k_z/\omega_0+\Delta\omega$, one finds $\omega^2-c^2k^2\simeq -c^2k_y^2-c^2k_z^2(1-c^2k_0^2/\omega_0^2)$. We have neglected the terms $(\Delta\omega)^2$ and $ck_z(\Delta\omega)$, which are much smaller than $c^2k_y^2$ (this can be verified a posteriori from Eqs. \ref{eq:g1}-\ref{eq:k1z}). Since $1-c^2k_0^2/\omega_0^2\simeq\omega_{\rm P}^2/\omega_0^2$, one finds $\omega^2 - c^2k^2\simeq -c^2k_y^2 -c^2k_z^2\omega_{\rm P}^2/\omega_0^2$.

Finally, we need to approximate $Q$. We discuss the two cases $\omega^2\gg c_{\rm s}^2k^2$ and $\omega^2\ll c_{\rm s}^2k^2$ below.\footnote{As discussed by \citet[][]{Ghosh+2021}, when $\omega^2\gg c_{\rm s}^2k^2$ the ponderomotive force is balanced by the electron inertia. The reason is that one may neglect $c_{\rm s}^2\nabla\delta N/N_0$ with respect to $\pa\delta{\bf V}/\pa t$ in Eq. \eqref{eq:p3}. When $\omega^2\ll c_{\rm s}^2k^2$,  the thermal pressure and the ponderomotive force balance each other since one may neglect $\pa\delta{\bf V}/\pa t$ with respect to $c_{\rm s}^2\nabla\delta N/N_0$ in Eq. \eqref{eq:p3}.} When $\omega^2\gg c_{\rm s}^2k^2$, one finds $Q\simeq c^2k^2/\omega^2$. When $c^2k_z^2\ll (\Delta\omega)^2$, one can approximate $Q\simeq c^2k_y^2/(\Delta\omega)^2$ and $\omega^2 - c^2k^2\simeq -c^2k_y^2$. Then the dispersion relation can be approximated as
\begin{equation}
\label{eq:fil1}
4\omega_0^2\left(\frac{\Delta\omega}{ck_y}\right)^4 -c^2k_y^2\left(\frac{\Delta\omega}{ck_y}\right)^2-\frac{1}{2}a_0^2\omega_{\rm P}^2 =0 \;.
\end{equation}
The wave number of the most unstable modes is $ck_y \gg \sqrt{a_0\omega_0\omega_{\rm P}}$, and the corresponding growth rate is determined by $(\Delta\omega)^2 = -a_0^2\omega_{\rm P}^2/2$. Since $\Delta\omega$ is purely imaginary, the perturbation moves along $z$ with the group velocity of the pump wave, $c^2k_0/\omega_0$. The conditions $c_{\rm s}^2k^2\ll\omega^2$ and $c^2k_z^2\ll (\Delta\omega)^2$ give $c_{\rm s}k_y\ll a_0\omega_{\rm P}$ and $ck_z\ll a_0\omega_{\rm P}$. Following the procedure that we used to derive Eq. \eqref{eq:fil1}, one sees that the instability does not develop for $c^2k_z^2\gg (\Delta\omega)^2$, which gives $Q\simeq k^2/k_z^2$.

When $\omega^2\ll c_{\rm s}^2k^2$, one finds $Q\simeq -c^2/c_{\rm s}^2$. Then the dispersion relation can be approximated as
\begin{equation}
\label{eq:fil2}
4\omega_0^2\left(\Delta\omega\right)^2 = c^2k_y^2 \left(c^2k_y^2 -\frac{1}{2}a_0^2\frac{c^2}{c_{\rm s}^2} \omega_{\rm P}^2\right) \;.
\end{equation}
The maximum growth rate of the instability is found when $c^2k_y^2=a_0^2c^2\omega_{\rm P}^2/4 c_{\rm s}^2$, which gives $(\Delta\omega)^2 = -a_0^4c^4\omega_{\rm P}^4/64\omega_0^2c_{\rm s}^4$. The condition $\omega^2\ll c_{\rm s}^2k^2$ requires $(\Delta\omega)^2\ll c_{\rm s}^2k_y^2$ and $c^2k_z^2\ll c_{\rm s}^2k_y^2$, which give $a_0\ll c_{\rm s}^2\omega_0/\omega_{\rm P}c^2$ and $ck_z\ll a_0\omega_{\rm P}$ respectively.

The instability is robust since it can be excited over a broad range of wave numbers. As we show in Appendix \ref{sec:appA}, the instability develops for all wave numbers $k_y>0$ in a cold plasma where the dispersion relation is given by Eq. \eqref{eq:fil1}. The instability develops for $0<k_y<a_0\omega_{\rm P}/\sqrt{2}c_{\rm s}$ in a warm plasma where the dispersion relation is given by Eq. \eqref{eq:fil2}.

We summarise our results in Eq. \eqref{eq:g1}-\eqref{eq:k1z} below. The growth rate can be estimated as
\begin{align}
\label{eq:g1}
\Gamma & \simeq \frac{1}{\sqrt{2}}a_0\omega_{\rm P} & \left(a_0\gg \beta_{\rm s}^2\omega_0/\omega_{\rm P}\right) \\
\label{eq:g1bis}
\Gamma & \simeq \frac{1}{8}a_0^2\beta_{\rm s}^{-2}\frac{\omega_{\rm P}^2}{\omega_0} & \left(a_0\ll \beta_{\rm s}^2\omega_0/\omega_{\rm P}\right)
\end{align}
where $\beta_{\rm s}=c_{\rm s}/c$. The most unstable transverse wave number can be estimated as
\begin{align}
\label{eq:k1y}
\frac{\sqrt{a_0\omega_0\omega_{\rm P}}}{c} \ll k_y & \ll a_0 \beta_{\rm s}^{-1}\frac{\omega_{\rm P}}{c}  & \left(a_0\gg \beta_{\rm s}^2\omega_0/\omega_{\rm P}\right) \\
\label{eq:k1ybis}
k_y & \simeq \frac{1}{2}a_0\beta_{\rm s}^{-1}\frac{\omega_{\rm P}}{c}  & \left(a_0\ll \beta_{\rm s}^2\omega_0/\omega_{\rm P}\right)
\end{align}
When $a_0\gg \beta_{\rm s}^2\omega_0/\omega_{\rm P}$, the growth rate remains same order of the maximal one for $\sqrt{a_0\omega_0\omega_{\rm P}} \ll ck_y\ll a_0\beta_{\rm s}^{-1}\omega_{\rm P}$. The most unstable longitudinal wave number can be estimated as
\begin{equation}
\label{eq:k1z}
k_z \ll a_0\frac{\omega_{\rm P}}{c}
\end{equation}
The growth rate remains same order of the maximal one for $k_z\ll a_0\omega_{\rm P}/c$, while there is no instability for $k_z\gg a_0\omega_{\rm P}/c$. Since $k_y\gg k_z$, the modulations are elongated in the direction of propagation of the electromagnetic pump wave (the instability breaks a wave packet into longitudinal filaments). Since $\omega^2\sim c^2k_z^2+\Gamma^2\sim a_0^2\omega_{\rm P}^2$, the condition $\omega^2\gg\omega_{\rm L}^2$ could be satisfied only for a weak magnetisation $\sigma_{\rm w}\ll a_0^2\lesssim 1$.

The modes described by Eqs. \eqref{eq:g1}-\eqref{eq:k1z} also exist in unmagnetised electron-ion plasmas \citep[e.g.][]{Drake+1974, Sobacchi+2021}, with the only difference that the ion plasma frequency replaces the electron plasma frequency. Our results are consistent with those of \citet[][]{Ghosh+2021}, who studied the filamentation of electromagnetic waves in unmagnetised pair plasmas.

\begin{figure*}{\vspace{3mm}} 
\centering
\includegraphics[width=0.99\textwidth]{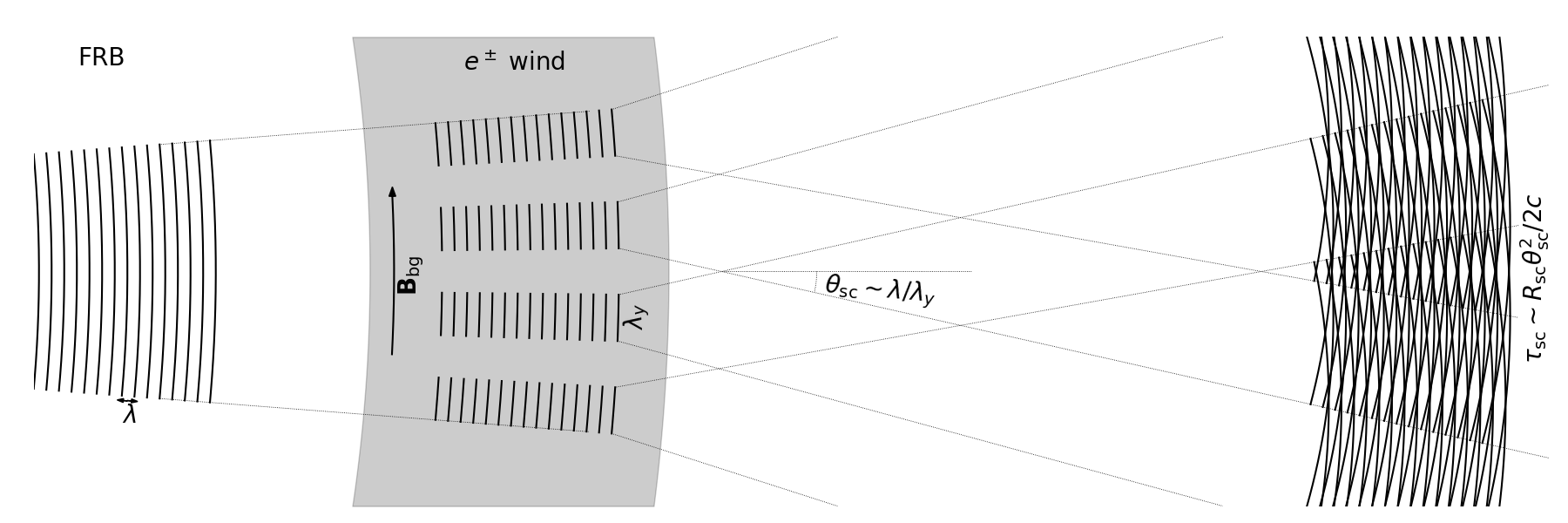}
\caption{Sketch of the effect of the filamentation instability on the FRB. The magnetar wind  (grey region) is pictured as a plasma slab of radius $R_{\rm sc}$ and thickness $\Delta R\sim R_{\rm sc}$ (see Eqs. \ref{eq:Rsc}, \ref{eq:R1}, and \ref{eq:R2}). The FRB electromagnetic wave (black lines) is broken into sheets of transverse size $\lambda_y$, perpendicular to the direction of the wind magnetic field, ${\bf B}_{\rm bg}$. As the FRB front expands, the radiation sheets are scattered due to diffraction by an angle $\theta_{\rm sc}\sim\lambda/\lambda_y$, where $\lambda$ is the FRB wavelength in the observer's frame. The corresponding scattering time is $\tau_{\rm sc}\sim R_{\rm sc}\theta_{\rm sc}^2/2c$.
}
\label{figure1}
\end{figure*}

\subsubsection{Case $\omega^2\ll\omega_{\rm L}^2$, with $B_{{\rm bg,}x}=0$}
\label{sec:fast}

Since $\omega^2\gg\omega_{\rm L}^2$ could be satisfied only for a weak magnetisation $\sigma_{\rm w}\ll a_0^2\lesssim 1$, one should study the case $\omega^2\ll\omega_{\rm L}^2$.

When $B_{{\rm bg,}x}=0$, the ponderomotive force is nearly parallel to the background magnetic field. Indeed, one finds $\cos^2\theta=k_y^2/(k_y^2+k_z^2)$, which gives $\theta\sim 0$ for $k_z\ll k_y$. Then one can approximate $Q\simeq c^2k^2/(\omega^2-c_{\rm s}^2k^2)$, which is the same as in the weakly magnetised case discussed in the previous section. The most unstable wave number and the growth rate are given by Eqs. \eqref{eq:g1}-\eqref{eq:k1z}. These results are summarised in Table \ref{table1}. Since $\omega^2\sim c^2k_z^2+\Gamma^2\sim a_0^2\omega_{\rm P}^2$, the condition $\omega^2\ll\omega_{\rm L}^2$ is satisfied in a magnetically-dominated plasma.

We remark that the wave number and the growth rate of the most unstable modes are the same as in the weakly magnetised case. The reason is that the particles can move freely along the background magnetic field under the effect of the ponderomotive force when $\theta\sim 0$.

\subsubsection{Case $\omega^2\ll\omega_{\rm L}^2$, with $B_{{\rm bg,}y}=0$}
\label{sec:slow}

When $B_{{\rm bg,}y}=0$, the ponderomotive force is perpendicular to the background magnetic field. For $\cos\theta=0$, one finds $Q\ll 1$ because $\omega_{\rm L}^2$ appears only in the denominator of $Q$, and $\omega^2$, $c^2k^2$, $\omega_{\rm P}^2$ are much smaller than $\omega_{\rm L}^2$ in magnetically-dominated plasmas. For $Q\ll 1$ the dispersion relation can be approximated as
\begin{equation}
\label{eq:fil3}
4\omega_0^2\left(\Delta\omega\right)^2=\left(c^2k_y^2+\frac{\omega_{\rm P}^2}{\omega_0^2}c^2k_z^2\right)\left(c^2k_y^2+\frac{\omega_{\rm P}^2}{\omega_0^2}c^2k_z^2-\frac{1}{2}a_0^2\omega_{\rm P}^2\right)\;.
\end{equation}
The condition $\omega^2\ll\omega_{\rm L}^2$ gives $c^2k^2_z\ll\omega_{\rm L}^2$. Since $(\omega_{\rm P}^2/\omega_0^2)c^2k_z^2\ll\omega_{\rm P}^2\omega_{\rm L}^2/\omega_0^2\ll a_0^2\omega_{\rm P}^2$, the terms proportional to $c^2k_z^2$ can be neglected in Eq. \eqref{eq:fil3}. The wave number of the most unstable modes is $c^2k^2_y = a_0^2\omega_{\rm P}^2/4$, and the corresponding growth rate is determined by $(\Delta\omega)^2 = -a_0^4\omega_{\rm P}^4/64\omega_0^2$. We conclude that the growth rate can be estimated as
\begin{equation}
\label{eq:g2}
\Gamma \simeq \frac{1}{8}a_0^2\frac{\omega_{\rm P}^2}{\omega_0}
\end{equation}
and the most unstable wave number can be estimated as
\begin{align}
\label{eq:k2y}
k_y & \simeq \frac{1}{2}a_0\frac{\omega_{\rm P}}{c} \\
\label{eq:k2z}
k_z & \ll \frac{\omega_{\rm L}}{c}\;.
\end{align}
These results are summarised in Table \ref{table2}. Comparing Eqs. \eqref{eq:g1}-\eqref{eq:g1bis} and \eqref{eq:g2}, one sees that the growth rate is faster when the ponderomotive force is nearly parallel to the direction of the background magnetic field. This may explain the formation of density sheets nearly perpendicular to the pre-shock magnetic field in three-dimensional simulations of the relativistic magnetised shocks \citep[][]{Sironi+2021}.

In the modes described by Eqs. \eqref{eq:g2}-\eqref{eq:k2z}, the dominant non-linear effect is the relativistic correction to the electron motion. The effect of the ponderomotive force is suppressed since the particles cannot move in the direction perpendicular to the magnetic field, and the instability develops at nearly constant electron density. The same modes also exist in unmagnetised electron-ion plasmas \citep[e.g.][]{Max+1974, Sobacchi+2021}, where the ponderomotive force can be suppressed due to the inertia of the ions.

In magnetically-dominated plasmas the particle distribution could be anisotropic. Then the thermal velocity $c_{\rm s}$ may be different along the magnetic field lines and in the perpendicular direction. Since the unstable wave numbers and the growth rate are independent of $c_{\rm s}$ when the ponderomotive force is perpendicular to the background magnetic field, our results depend only on the value of the thermal velocity along the field lines.

\section{Scattering of Fast Radio Bursts}
\label{sec:scattering}

We apply these results to the propagation of a FRB through the magnetically-dominated magnetar wind. The observed duration of the FRB is $T\sim 1{\rm\; ms}$. Since the FRB light curve is typically variable, we consider the possibility that the burst is made of pulses with duration $\tau<T$, during which the radiation intensity remains constant. We do not assume any specific emission mechanism.

The wind may be thought of as a sequence of plasma slabs of thickness $\Delta R\sim R$ with a decreasing plasma density $n(R)$. The instability develops when:
\begin{enumerate}
\item The longitudinal wavelength of the unstable modes, $\lambda_z\simeq 2\pi/k_z$, is shorter than the length of the pulse in the wind frame, $\ell\simeq 2\gamma_{\rm w}c\tau$.
\item The timescale on which the instability grows, $t_{\rm gr}\simeq 10/\Gamma$, is shorter than the expansion time of the wave front in the wind frame, $t_{\rm exp}\simeq R/\gamma_{\rm w}c$.
\end{enumerate}
The conditions (i) and (ii) are satisfied for $R\lesssim R_1$ and $R\lesssim R_2$ respectively. The values of $R_1$ and $R_2$ depend on the thermal velocity, and we calculate them below in the relevant cases.

The instability breaks the wave packet into sheets of radiation perpendicular to the direction of the wind magnetic field. We estimate the transverse size of the radiation sheets as $\lambda_y\simeq 2\pi/k_y$, where $k_y(R)$ is the wave number of the most unstable modes.\footnote{We remark that our estimate relies on an extrapolation of the results of the linear stability analysis, and further investigation is required to understand how the instability saturates.} At radii $R\ll\min[R_1,R_2]$, the most unstable transverse wavelength slowly increases with the radius, namely ${\rm d}\lambda_y/{\rm d}R\ll \theta_{\rm sc}(R)$, where $\theta_{\rm sc}(R)\sim\lambda/\lambda_y$ is the scattering angle at the radius $R$ ($\theta_{\rm sc}$ is measured in the observer's frame), and $\lambda=\pi c/\gamma_{\rm w}\omega_0$ is the FRB wavelength in the observer's frame.\footnote{In a linear wave, transverse modulations of the wave intensity with a scale $\lambda_y$ result in the deflection of the wave through an angle $\theta_{\rm sc}\sim\lambda/\lambda_y$, which may be thought of as diffraction scattering. Non-linear effects prevent diffraction from occurring at radii $R\ll\max[R_1,R_2]$.} Then the transverse scale of the sheets is gradually adjusted to $\lambda_y$. Scattering occurs at a large radius
\begin{equation}
\label{eq:Rsc}
R_{\rm sc}=f\min\left[R_1,R_2\right]\;,
\end{equation}
where $f\sim 1$ is a numerical factor, since the instability no longer develops for $R\gtrsim R_{\rm sc}$. The corresponding scattering time is $\tau_{\rm sc}\sim R_{\rm sc}\theta_{\rm sc}^2/2c$. The outlined scenario is sketched in Figure \ref{figure1}.

Different frequency components of the same burst have different scattering times. Since filamentation is a non-linear process, the transverse scale of the sheets, $\lambda_y$, depends on the power-weighted frequency of the burst. On the other hand, low frequency components are more diffracted. The scattering angle is $\theta_{\rm sc}\sim \lambda/\lambda_y\propto\nu^{-1}$, and the corresponding scattering time is $\tau_{\rm sc}\sim R_{\rm sc}\theta_{\rm sc}^2/2c\propto \nu^{-2}$.

We consider the case when the transverse component of the perturbation wave vector is parallel to the wind magnetic field (see Section \ref{sec:fast} and Table \ref{table1}), which gives the largest growth rate of the instability. First we discuss the case of a warm plasma, and then the case of a cold plasma.\footnote{The magnetar wind cools down radiatively and adiabatically. On the other hand, the wind could be heated by magnetic reconnection \citep[e.g.][]{LyubarskyKirk2001}, and by internal shocks \citep[e.g.][]{Beloborodov2020}. We consider the possibility that these processes keep the plasma warm.}

\subsection{Warm magnetar wind}

We start considering the case of a warm plasma with $a_0\ll\beta_{\rm s}^2\omega_0/\omega_{\rm P}$, which is satisfied for
\begin{align}
\beta_{\rm s} & \gg 2\times 10^{-3} L_{42}^{1/4}\mu_{33}^{1/2}P_0^{-1}\nu_9^{-1}R_{15}^{-1}\sigma_{\rm w}^{-1/4} \sim \nonumber\\
& \sim 8\times 10^{-4} L_{42}^{1/4} \dot{N}_{42}^{1/4} \nu_9^{-1} R_{15}^{-1} \gamma_{\rm w}^{1/4} \;.
\end{align}
We have expressed $\beta_{\rm s}$ as a function of the rate of particle outflow in the wind, $\dot{N}\sim L_{\rm w}/\gamma_{\rm w}\sigma_{\rm w}mc^2$, where $L_{\rm w}\sim\mu^2(2\pi/P)^4/c^3$ is the luminosity of the wind ($\mu$ and $P$ are the magnetic dipole moment and the rotational period of the magnetar, $\gamma_{\rm w}$ and $\sigma_{\rm w}$ are the Lorentz factor and the magnetisation of the wind). For our fiducial parameters, we find $\dot{N}\sim 7\times 10^{43}\mu_{33}^2P_0^{-4}\gamma_{\rm w}^{-1}\sigma_{\rm w}^{-1}{\rm\; s^{-1}}$. We have defined $\dot{N}_{42}\equiv\dot{N}/10^{42}{\rm\; s^{-1}}$, which is the appropriate normalisation for values of $\gamma_{\rm w}\sigma_{\rm w}$ of the order of a few tens, as we find below.

The wave numbers and the growth rate of the most unstable mode are $ck_y\simeq a_0\beta_{\rm s}^{-1} \omega_{\rm P}/2$, $ck_z\lesssim a_0\omega_{\rm P}$, and $\Gamma\simeq a_0^2\beta_{\rm s}^{-2}\omega_{\rm P}^2/8\omega_0$. Then one finds
\begin{align}
\nonumber
R_1 & \sim 2\times 10^{15} L_{42}^{1/4}\mu_{33}^{1/2}P_0^{-1}\nu_9^{-1/2}\tau_{-3}^{1/2}\sigma_{\rm w}^{-1/4}{\rm\; cm} \sim \\
\label{eq:R1}
& \sim 8\times 10^{14} L_{42}^{1/4}\dot{N}_{42}^{1/4}\nu_9^{-1/2}\tau_{-3}^{1/2}\gamma_{\rm w}^{1/4}{\rm\; cm}\\
\nonumber
R_2 & \sim 4\times 10^{15} L_{42}^{1/3}\mu_{33}^{2/3}P_0^{-4/3}\nu_9^{-1}\beta_{\rm s}^{-2/3}\gamma_{\rm w}^{-2/3}\sigma_{\rm w}^{-1/3} {\rm\; cm}\sim \\
\label{eq:R2}
& \sim 9\times 10^{14} L_{42}^{1/3}\dot{N}_{42}^{1/3}\nu_9^{-1}\beta_{\rm s}^{-2/3}\gamma_{\rm w}^{-1/3} {\rm\; cm}\;,
\end{align}
where we have defined $\tau_{-3}\equiv\tau/1{\rm\; ms}$. The conditions $\lambda_z\lesssim\ell$ and $t_{\rm gr}\lesssim t_{\rm exp}$ are satisfied at radii $R\lesssim R_1$ and $R\lesssim R_2$ respectively.

The value of the scattering time depends on the Lorentz factor of the wind. The critical wind Lorentz factor that gives $R_1=R_2$ is
\begin{align}
\nonumber
\gamma_{\rm cr} & \sim 2\; L_{42}^{1/8}\mu_{33}^{1/4}P_0^{-1/2} \nu_9^{-3/4} \tau_{-3}^{-3/4}\beta_{\rm s}^{-1} \sigma_{\rm w}^{-1/8} \sim \\
& \sim L_{42}^{1/7}\dot{N}_{42}^{1/7}\nu_9^{-6/7}\tau_{-3}^{-6/7} \beta_{\rm s}^{-8/7} \;.
\end{align}
Note that $\gamma_{\rm cr}\gg 1$ for $\beta_{\rm s}\ll 1$. When $\gamma_{\rm w}\lesssim\gamma_{\rm cr}$, one finds $R_{\rm sc}=fR_1$, and the scattering time is
\begin{align}
\nonumber
\tau_{\rm sc} & \sim 3\; L_{42}^{1/4}\mu_{33}^{1/2}P_0^{-1}\nu_9^{-5/2}\tau_{-3}^{-3/2}\beta_{\rm s}^{-2} \gamma_{\rm w}^{-2}\sigma_{\rm w}^{-1/4} f^{-3} {\rm\; ns} \sim \\
\label{eq:tausc1}
& \sim 0.9\; L_{42}^{1/4}\dot{N}_{42}^{1/4}\nu_9^{-5/2}\tau_{-3}^{-3/2}\beta_{\rm s}^{-2} \gamma_{\rm w}^{-7/4}f^{-3} {\rm\; ns} \;.
\end{align}
When $\gamma_{\rm w}\gtrsim\gamma_{\rm cr}$, one finds $R_{\rm sc}=fR_2$, and the scattering time is
\begin{equation}
\label{eq:tausc2}
\tau_{\rm sc}\sim 0.8\;\nu_9^{-1} f^{-3} {\rm\; ns} \;.
\end{equation}

Scattering in a warm wind produces a frequency modulation with a large bandwidth, $\Delta\nu\sim 1/\tau_{\rm sc}\gtrsim 100{\rm\; MHz}$ (see Eqs. \ref{eq:tausc1} and \ref{eq:tausc2}). Such broadband frequency modulations are observed in FRBs \citep[e.g.][]{Shannon+2018, Hessels+2019, Nimmo+2021}. The bandwidth increases with the burst frequency (Eqs. \ref{eq:tausc1} and \ref{eq:tausc2} give $\Delta\nu\propto\nu^\beta$ with $\beta\sim 1-2.5$), consistent with observations of FRB 121102 \citep[][]{Hessels+2019}.

When the transverse component of the perturbation wave vector is perpendicular to the wind magnetic field (see Section \ref{sec:slow} and Table \ref{table2}), the unstable modes have $ck_y\simeq a_0\omega_{\rm P}/2$, $ck_z\lesssim \omega_{\rm L}$, and $\Gamma\simeq a_0^2\omega_{\rm P}^2/8\omega_0$. Since the condition $\lambda_z\lesssim\ell$ is easily satisfied, one finds $R_{\rm sc}= fR_2$. The scattering time is the same as in Eq. \eqref{eq:tausc2}, regardless of $\beta_{\rm s}$.

\subsection{Cold magnetar wind}

In a cold plasma with $a_0\gg\beta_{\rm s}^2\omega_0/\omega_{\rm P}$, we have $ck_z\lesssim a_0\omega_{\rm P}$ and $\Gamma\simeq a_0\omega_{\rm P}/\sqrt{2}$. Since the condition $t_{\rm gr}\lesssim t_{\rm exp}$ is easily satisfied, one finds $R_{\rm sc}= fR_1$. The growth rate remains same order of the maximal one for the transverse wave numbers $\sqrt{a_0 \omega_0 \omega_{\rm P}} \lesssim ck_y\lesssim a_0\beta_{\rm s}^{-1}\omega_{\rm P}$. To determine the dominant transverse scale of the radiation sheets, $\lambda_y$, one should study how the instability saturates for various wave numbers, which is out of the scope of the paper. Nevertheless, our results can be used to place a lower limit on the Lorentz factor of the magnetar wind. Since $\lambda_y\lesssim2\pi c/\sqrt{a_0\omega_0\omega_{\rm P}}$, one finds a lower limit for the scattering time that is independent of $\beta_{\rm s}$,
\begin{align}
\nonumber
\tau_{\rm sc} & \gtrsim 10\; L_{42}^{1/4}\mu_{33}^{1/2}P_0^{-1}\nu_9^{-1/2}\tau_{-3}^{-1/2}\gamma_{\rm w}^{-2}\sigma_{\rm w}^{-1/4} f^{-1} {\rm\; ms} \sim \\
\label{eq:tausc3}
& \sim 4\; L_{42}^{1/4}\dot{N}_{42}^{1/4}\nu_9^{-1/2}\tau_{-3}^{-1/2}\gamma_{\rm w}^{-7/4}f^{-1} {\rm\; ms} \;.
\end{align}
We remark that $\nu_9$ corresponds to the power-weighted frequency of the burst. As discussed above, the frequency components of one burst have different scattering times, with $\tau_{\rm sc}\propto\nu^{-2}$.

Interestingly, the frequency-dependent broadening of the brightest pulse from FRB 181112 is consistent with $\tau_{\rm sc}\propto\nu^{-2}$ \citep[][]{Cho+2020}. The rise time of the pulse is $\tau\sim 15 {\rm\;\mu s}$, and the scattering time is $\tau_{\rm sc}\sim 25 {\rm\;\mu s}$. The observed scattering can be an effect of propagation through a cold magnetar wind. Substituting $\tau\sim 15 {\rm\;\mu s}$ and $\tau_{\rm sc}\sim 25 {\rm\;\mu s}$ into Eq. \eqref{eq:tausc3}, one can estimate the wind Lorentz factor,
\begin{align}
\nonumber
\gamma_{\rm w} & \gtrsim 60\; L_{42}^{1/8}\mu_{33}^{1/4}P_0^{-1/2}\nu_9^{-1/4} \sigma_{\rm w}^{-1/8} f^{-1/2} \sim \\
& \sim 50\; L_{42}^{1/7}\dot{N}_{42}^{1/7}\nu_9^{-2/7}f^{-4/7}\;.
\end{align}
This Lorentz factor is not far from $\gamma_{\rm w} \sim 10-30$ estimated for magnetar winds \citep[][]{Beloborodov2020}.

\section{Conclusions}
\label{sec:conclusions}

We have studied the modulation/filamentation instabilities of FRBs propagating in a magnetar wind. We have modelled the wind as a magnetically-dominated pair plasma. We have focused on the regime of sub-relativistic electron motion, i.e. dimensionless wave strength parameter $a_0\lesssim 1$.

The instability modulates the intensity of the radio wave, producing radiation sheets perpendicular to the direction of the wind magnetic field. As the FRB front expands outside the radius where the instability ends, the radiation sheets are diffracted, effectively spreading the arrival time of the FRB wave. The imprint of the scattering on the time-frequency structure of FRBs depends on the properties of the wind.

In a cold wind with $\beta_{\rm s}\ll 10^{-3}$ ($\beta_{\rm s}=c_{\rm s}/c$ is the ratio of the thermal velocity along the magnetic field lines and the speed of light), the typical FRB scattering time is $\tau_{\rm sc}\sim{\rm\mu s-ms}$ at the frequency $\nu\sim 1{\rm\; GHz}$. Low frequencies have longer scattering times, with $\tau_{\rm sc}\propto\nu^{-2}$. Such frequency-dependent broadening has been observed in the brightest pulse of FRB 181112 \citep[][]{Cho+2020}. From the rise and scattering timescales of the pulse, we estimate the wind Lorentz factor, $\gamma_{\rm w}\gtrsim 50$. Within the accuracy of this estimate (a factor of a few), $\gamma_{\rm w}$ is consistent with theoretical expectations for magnetar winds \citep[][]{Beloborodov2020}.

In a warm wind with $\beta_{\rm s}\gg 10^{-3}$, the FRB scattering time can approach $\tau_{\rm sc}\sim{\rm ns}$. Then scattering produces a frequency modulation of the observed intensity with a large bandwidth, $\Delta\nu\sim 1/\tau_{\rm sc}\gtrsim 100{\rm\; MHz}$. The modulation bandwidth increases with the burst frequency. Broadband frequency modulations observed in FRBs \citep[e.g.][]{Shannon+2018, Hessels+2019, Nimmo+2021} could be due to scattering in a warm magnetar wind.

\section*{Acknowledgements}

We thank the anonymous referee for constructive comments and suggestions that improved the paper. We acknowledge fruitful discussions with Masanori Iwamoto. YL acknowledges support from the Israeli Science Foundation grant 2067/19. AMB acknowledges support from the Simons Foundation grant \#446228, the Humboldt Foundation, and NSF AST-2009453. LS acknowledges support from the Sloan Fellowship, the Cottrell Scholars Award, NASA 80NSSC20K1556, NASA 80NSSC18K1104, and NSF AST-1716567.

\section*{Data availability}

No new data were generated or analysed in support of this research.

\def\aap{A\&A}\def\aj{AJ}\def\apj{ApJ}\def\apjl{ApJ}\def\mnras{MNRAS}\def\prl{Phys. Rev. Lett.}
\def\araa{ARA\&A}\def\physrep{PhR}\def\sovast{Sov. Astron.}\def\nar{NewAR}\def\pasa{PASA}
\def\aapr{Astronomy \& Astrophysics Review}\def\apjs{ApJS}\def\nat{Nature}\def\na{New Astron.}
\def\prd{Phys. Rev. D}\def\pre{Phys. Rev. E}\def\pasp{PASP}\def\apss{ApSS}
\bibliographystyle{mn2e}
\bibliography{2d}

\appendix
\section{Solution of Eqs. (35), (36), (42)}
\label{sec:appA}

\begin{figure}{\vspace{3mm}} 
\centering
\includegraphics[width=0.45\textwidth]{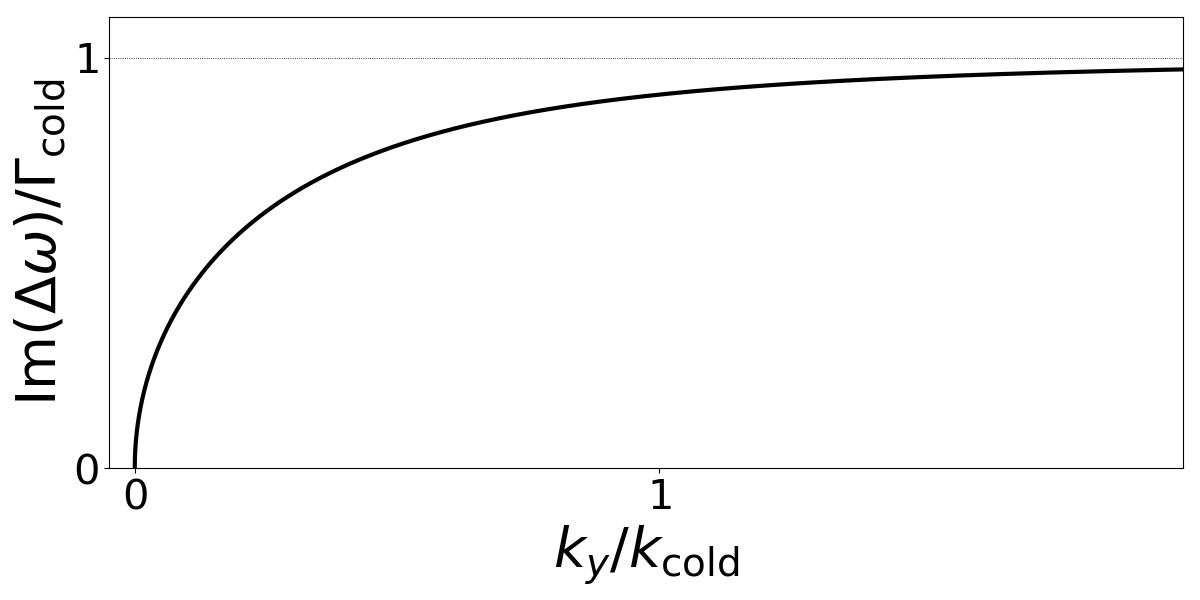}
\includegraphics[width=0.45\textwidth]{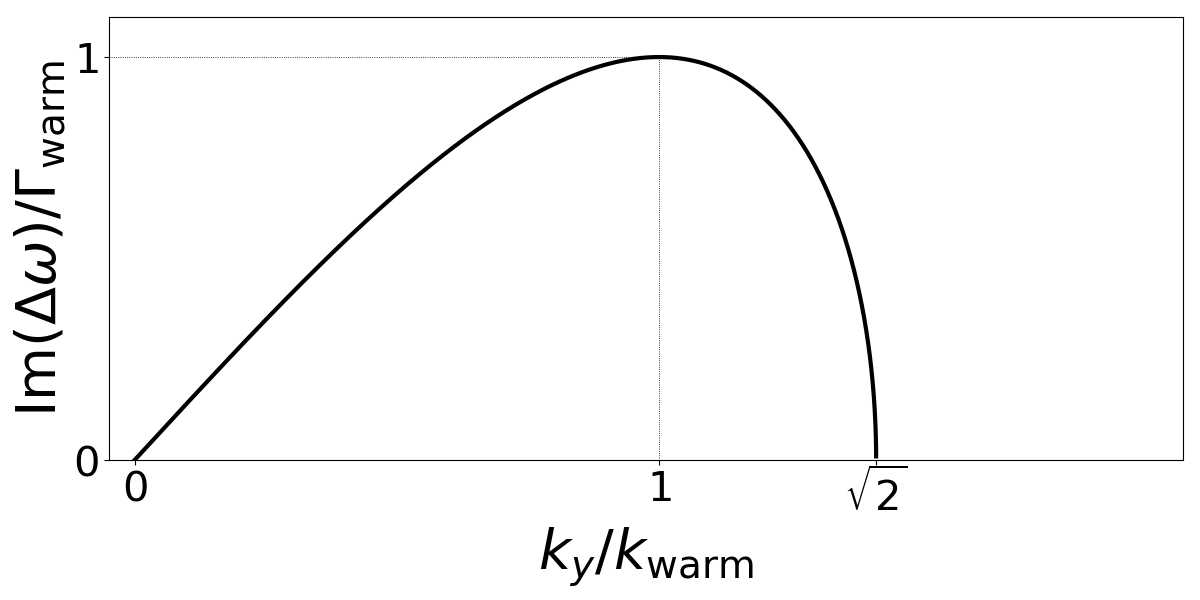}
\caption{Imaginary part of $\Delta\omega$ as a function of $k_y$ in a cold plasma (top panel) and in a warm plasma (bottom panel). The ponderomotive force is nearly parallel to the background magnetic field.
}
\label{fig:DR}
\end{figure}

We start with the case when the ponderomotive force is nearly parallel to the background magnetic field. In a cold plasma, the dispersion relation is given by Eq. \eqref{eq:fil1}. The growth rate of the exponentially growing solution can be presented as
\begin{equation}
\left(\Delta\omega\right)^2=-\frac{2\Gamma_{\rm cold}^2}{1+\sqrt{1+\frac{k_{\rm cold}^2}{k_y^2}}} \;,
\end{equation}
where $\Gamma_{\rm cold}=a_0\omega_{\rm P}/2$ and $ck_{\rm cold}=\sqrt[4]{8}\sqrt{a_0\omega_0\omega_{\rm P}}$. The maximal growth rate $\Gamma_{\rm cold}$ is achieved for $k_y\gg k_{\rm cold}$. In the top panel of Figure \ref{fig:DR} we plot ${\rm Im}(\Delta\omega)/\Gamma_{\rm cold}$ as a function of $k_y/k_{\rm cold}$. The instability develops for all $k_y$.

In a warm plasma, the dispersion relation is given by Eq. \eqref{eq:fil2}. The solution can be presented as
\begin{equation}
\left(\Delta\omega\right)^2= -\Gamma_{\rm warm}^2\left[1-\left(1-\frac{k_y^2}{k_{\rm warm}^2}\right)^2\right] \;,
\end{equation}
where $\Gamma_{\rm warm}=a_0^2\beta_{\rm s}^{-2}\omega_{\rm P}^2/8\omega_0$ and $ck_{\rm warm}=a_0\beta_{\rm s}^{-1}\omega_{\rm P}/2$. The maximal growth rate $\Gamma_{\rm warm}$ is achieved for $k_y= k_{\rm warm}$. In the bottom panel of Figure \ref{fig:DR} we plot ${\rm Im}(\Delta\omega)/\Gamma_{\rm warm}$ as a function of $k_y/k_{\rm warm}$. The instability develops for $k_y<\sqrt{2}k_{\rm warm}$.

When the ponderomotive force is perpendicular to the background magnetic field, the dispersion relation is given by Eq. \eqref{eq:fil3}. The terms proportional to $k_z^2$ are negligibly small. Then Eq. \eqref{eq:fil3} is identical to Eq. \eqref{eq:fil2} after the formal substitution $c_{\rm s}\to c$ in Eq. \eqref{eq:fil2}. The dependence of ${\rm Im}(\Delta\omega)$ on $k_y$ is analogous to the bottom panel of Figure \ref{fig:DR}.

\end{document}